\RequirePackage{etoolbox}
\patchcmd{\bibliographystyle}{#1}{sn-basic-unsort}{}{}
\documentclass[sn-basic, Numbered]{sn-jnl}

\usepackage{graphicx}%
\usepackage{multirow}%
\usepackage{amsmath,amssymb,amsfonts}%
\usepackage{amsthm}%
\usepackage{mathrsfs}%
\usepackage[title]{appendix}%
\usepackage{xcolor}%
\usepackage{textcomp}%
\usepackage{manyfoot}%
\usepackage{booktabs}%
\usepackage{algorithm}%
\usepackage{algorithmicx}%
\usepackage{algpseudocode}%
\usepackage{listings}%
\usepackage{ltablex}
\usepackage{lscape}
\usepackage{setspace}
\usepackage{array}
\newcolumntype{P}[1]{>{\centering\arraybackslash}p{#1}}
\newcolumntype{L}[1]{>{\raggedright\arraybackslash}p{#1}}
\newcommand{\tabularxcaption}[2]{%
    \begingroup
    \setbox\tabcapbox\vbox{\tablecaptionfont\raggedright%
    {\bfseries \mbox{#1}\nobreak}{\hskip2mm}\mbox{#2}\vphantom{y}\par\vskip\belowcaptionskip}%
    \box\tabcapbox%
    \endgroup
}
\usepackage{url}

\usepackage{geometry}
\theoremstyle{thmstyleone}%
\theoremstyle{thmstyletwo}%

\theoremstyle{thmstylethree}%

\begin{document}

\title[Ethical landscape of RAS]{The ethical landscape of robot-assisted surgery}
\subtitle{A systematic review}

\author*[1]{\fnm{Joschka} \sur{Haltaufderheide}}\email{joschka.haltaufderheide@uni-potsdam.de}
\author[2]{\fnm{Stefanie} \sur{Pfisterer-Heise}}
\author[2]{\fnm{Dawid} \sur{Pieper}}
\author[1]{\fnm{Robert} \sur{Ranisch}}

\affil*[1]{\orgdiv{Juniorprofessorship for Medical Ethics with a focus on Digitization}, \orgname{Faculty for Health Sciences Brandenburg, University of Potsdam}, \orgaddress{\street{Am M\"uhlenberg 9}, \city{Potsdam}, \postcode{14476}, \state{Brandenburg}, \country{Germany}}}
\affil[2]{\orgdiv{Institute for Health Services and Health Systems Research}, \orgname{Center for Health Services Research Brandenburg, Faculty of Health Sciences Brandenburg, Brandenburg Medical School Theodor Fontane (MHB)}}


 \abstract{\textbf{Background:} Robot-assisted surgery has been widely adopted in recent years. However, compared to other health technologies operating in close proximity to patients in a vulnerable state, ethical issues of robot-assisted surgery have received less attention. Against the background of increasing automation that are expected to raise new ethical issues, this systematic review aims to map the state of the ethical debate in this field.\\
\textbf{Methods:} A protocol was registered in the international prospective register of systematic reviews (PROSPERO CRD42023397951). Medline via PubMed, EMBASE, CINHAL, Philosophers’ Index, IEEE Xplorer, Web of Science (Core Collection), Scopus and Google Scholar were searched in January 2023. Screening, extraction, and analysis were conducted independently by two authors. A qualitative narrative synthesis was performed.\\
\textbf{Results:} Out of 1,723 records, 66 records were included in the final 
dataset. Seven major strands of the ethical debate emerged during analysis. These 
include questions of harms and benefits, responsibility and control, 
professional-patient relationship, ethical issues in surgical training and learning, 
justice, translational questions, and economic considerations.\\
\textbf{Discussion:} The identified themes testify to a broad range of different and differing ethical issues requiring careful deliberation and integration into the surgical ethos. Looking forward, we argue that a different perspective in addressing robotic surgical devices might be helpful to consider upcoming challenges of automation.}




\maketitle
\section*{Background}

Robot-assisted surgery (RAS) has been widely adopted
in recent years \cite{Kawashima_etal_2019,Sheetz_etal_2020b,Rizzo_etal_2023,Mehta_etal_2022}.
RAS involves the use of sophisticated robotic platforms during invasive
surgical procedures to support or assist surgeons, enabling high accuracy
and precision \cite{Biswas_etal__2023}. Currently, around 40 different
robotic systems are commercially available \cite{Klodmann_etal_2021}
and are used in various surgical subfields. The most commonly used
is the Da Vinci surgical system, with roughly 6,000 devices in operation
worldwide that have, up to now, performed approximately 8.5 million
procedures \cite{Mayor_etal_2022}.

Although the origins of RAS can be traced back to the 1970s, with
commercially available devices emerging in the 1990s \cite{Kalan_etal_2010,Pugin_etal_2011},
it is particularly in the last decade that case volumes have increased
significantly, and RAS has diffused into general surgical practice
\cite{Sheetz_etal_2020b,Chung_etal_2021,Cepolina_Razzoli_2022}. Especially
in the USA, UK, Germany and Japan, growing case volumes have been
observed while RAS has become the standard of care in various procedures
\cite{Mayor_etal_2022}.

Compared to other health technologies that operate in close proximity
to vulnerable care recipients, the development and introduction of
RAS has received relatively little attention from both ethicists and
the public. Given its long history and its prevalence in surgical
practice, it is surprising that RAS has not been accompanied by a
broader debate. Ethical investigations have been dispersed across
various academic fields, including philosophy, medical ethics, philosophy
of technology, medicine, and computer sciences. Recent technological
advances have, however, sparked a renewed interest in these issues.
Commentators have highlighted their understanding of RAS as a potential
bridge technology, linking traditional laparoscopic approaches with
more advanced technical arrangements in surgery. This includes, for
example, using available data from the machines to evaluate and optimize
surgical workflows as well as combining RAS image data and artificial
intelligence (AI) to further the development of tissue recognition
and computer-based surgical automation \cite{Feuner_Park_2017} as
well as enabling more complex forms of interaction between surgeons
and machines \cite{Bergeles_Yang_2014}.

Increasing automation, use of more advanced devices, and evolving
patterns of interaction will inevitably raise a variety of new ethical
and social questions. This review is the first step in an ongoing
project that aims to anticipate and evaluate potential ethical issues
of cutting-edge and next-generation devices. It compiles the current
state of knowledge and understanding of ethical issues associated
with RAS, drawn from both experience and scholarly discussions on
current technology. The goal is to establish a comprehensive foundation
that maps ethical perspectives and value considerations, which can
then inform discussions around potential future developments. To this
end, we conducted a comprehensive systematic review addressing the
following research questions:
\begin{enumerate}
  \item What are the ethical issues surrounding the existing use of robot-assisted
surgery?
  \item How is RAS framed in discussions of ethical issues?
\end{enumerate}
This includes investigating the main themes of the ethical debate
and mapping how technical properties and perspectives on these devices
are connected to these themes. First, we provide an overview of our
methods for conducting this review. We then present the results of
our analysis, focusing on the main themes that can be identified in
the current literature. Finally, we discuss implications of these
findings.

\section*{Methods}

Systematic reviews of ethical issues differ from established approaches
in empirical sciences which cannot be transferred to ethical questions
directly \cite{Kahrass_etal__2023}. Following Droste et. al \cite{Droste_etal__2010},
we, hence used the adapted PIE scheme instead of PICO to determine
our key concepts. In reporting, we follow the guideline for reporting
of systematic reviews in ethics (RESERVE) \cite{Kahrass_etal__2023}. A review protocol was designed
and agreed upon by the authors. It was registered in the international
prospective register of systematic reviews (PROSPERO) \cite{Haltaufderheide_etal_2023}.
The goal was to analyze ethical issues emerging from practical experience,
such as those documented in empirical work, case reports or letters,
as well as from theoretical work from different scholarly perspectives. 

\subsection*{Inclusion and exclusion criteria}

We used two key concepts to define our inclusion criteria. The first,
RAS, was defined as surgical procedure with any surgical technology
that places a computer-aided electromechanical device in the path
between surgeon and patient \cite{TheSAGESMIRARoboticSurgeryConsensusGroup_etal_2008}.
We distinguished computer-enhanced devices from purely mechanical
manipulators based on their capacity to complex digital data processing.
A robotic surgical device, therefore, is any electrical system capable
of information processing that incorporates programmable actuator
mechanisms to facilitate in the placement of surgical instruments
within a patient.

We excluded all papers discussing exclusively hypothetical technical
arrangements such as fully autonomous RAS to avoid discussions not
grounded in experience and technical realities. Operationalizing this
criterion proved more difficult than expected. We planned to use a
taxonomy for levels of functional autonomy, following the categorization
of Yang et al. \cite{Yang_etal_2017}, subsequently refined by Lee
et al. \cite{Lee_etal_2024}. However, drawing on this approach turned
out to be difficult, as many authors in our records made only loose
reference to specific technical features, merely implied a certain
state of technology or discussed various scenarios. Consequently,
we resorted to an array of criteria as proxy indicators. We checked
whether and which devices were referenced by name, or whether reference
to real world cases was made. We hypothesized that this would keep
us in the range of commercially available and approved devices. Where
no device could be identified, the state of development (e.g. prototypes,
technical concept studies) or the reference to technical features
was unclear, we checked and discussed whether described or implied
features would lie within the autonomy levels 0 and 2 according to
Yang et al. In case of reference to various states of technology (e.g.
future and present) the work was included but only relevant parts
were extracted.

For our second key concept, ethical issues, we defined these as a
state in which moral implications of a given situation cannot be determined
without much reservation, where there is disagreement regarding to
the right course of action, or conflicting moral obligations or values
are present \cite{Schofield_etal_2021,Klingler_etal_2017}.
This can include a broad range of situations, such as unclear or undetermined
benefits, chances, risks, or harms; unclear, undetermined or conflicting
views about addressees of moral complaints or bearers of moral
significance as well as situations in which moral principles were evidently
overlooked or ignored. We determined an ethical issue to be present
whenever it was recognized as such in the literature or was identified
during analysis based on our criteria. 

We did not define any additional inclusion or exclusion criteria based
on, for example, specific patient groups, type of publication or type
of work presented .

\subsection*{Sources}

Database searches were conducted in January 2023. Databases included
Medline via PubMed, EMBASE, CINHAL, Philosopers’ Index, IEEE Xplorer,
Web of Science (Core Collection), Scopus, and Google Scholar. In Google
Scholar, we limited the search to the first 200 records. In addition,
we searched citations of included full-text articles, scanned conference
proceedings, and consulted experts from the field. Details on the
search strategy can be found in the protocol \cite{Haltaufderheide_etal_2023}.

\subsection*{Screening}

Two of the authors {[}J.H. and S.P-H{]} independently screened titles
and abstracts. Conflicts were resolved through discussion. In case
of prevailing disagreement in the title and abstract stage the decision
was postponed until the full-text screening stage. Full texts were
screened independently by the same authors. Disagreement was resolved
through discussion.

\subsection*{Extraction and analysis}

Data was subsequently extracted by the same authors using a self-designed extraction form.
Besides basic bibliographic information, the authors extracted data
on interventions and settings, information about devices referenced,
ethical values named or referenced, ethical arguments, key conclusions,
and further recommendations. For data extraction, these categories
were translated into a preliminary coding tree using MAXQDA 20. 

\subsection*{Synthesis}

A synthesized coding tree using the above-named categories was developed,
based on the first ten records in alphabetical order to ensure a shared
understanding of the coded features. These first ten records were
screened independently by two authors, and the results and coding
schemes were jointly discussed. Following the quality appraisal procedure
as outlined below, the first author coded all remaining papers and
at the same time categorized them to steer further extraction. The
second author, then independently coded all records of higher quality
and reviewed the rest for saturation with the help of a supervised assistant. Finally, all codes and categories
were synthesized.\begin{table}[hbt] 
\caption{\label{tab:assessment}Criteria for hybrid quality assessment}
\begin{tabularx}{\textwidth}{@{\extracolsep\fill}
L{\dimexpr1\textwidth-2\tabcolsep}}
\toprule
1. Weighing Quality in Extraction\\ 
\midrule
Is the publication peer-reviewed or otherwise quality controlled during the publication process?\\
Is the publication a comment, perspective, letter, viewpoint or editorial? \\
Does the content of the publication display material richness?\\
 \midrule
2. Weighing Quality in Synthesis\\ 
\midrule
Does the argument display a focused question?\\
Does the argument relate/engage with existing literature?\\
Does the argument denote a normative framework/principle?\\
Does the argument refer to empirical data in its descriptive claims?\\
Does the argument reflect on its assumed relationship of normative and empirical claims?\\
Does the argument come to a conclusion?\\
\midrule
3. Weighing Quality in Reporting\\ 
\midrule
Does the dataset/further literature display arguments or objections contrary to its position?\\
Does the dataset/literature indicate limits in generalizability or absoluteness of a claim?\\
Does the dataset/further literature display empirical evidence contrary to the displayed empirical claims?\\

\botrule
\end{tabularx}
\end{table}

\subsection*{Quality appraisal}

Currently no agreed upon criteria for the quality appraisal of normative
literature exist. However, we agree with Mertz et al. in that quality appraisal is an
important step in systematically reviewing literature \cite{Mertz_2019}.
Based on these considerations, we developed a hybrid strategy. Given
the aim to steer the process of reviewing towards higher level evidence
as well as the fact that content-related methods of appraisal require
to be included in the later stages of the process, we integrated quality
criteria in the extraction and analysis as well as the synthesis and
reporting stage giving increasing weight to content-related criteria
in the later steps \cite{Mertz_2019}. A first step of our quality
appraisal served to steer the extraction and analysis stage of our
review and was based on procedural criteria. We used this step to
inform the grouping of records in those which would be primarily extracted
and those which were used to check for saturation, hence giving weight
to those included studies that indicate to be epistemically more trustworthy
and display a certain material richness in content. In the second
step we used a modified version of the reporting criteria as proposed
by McCullough to appraise comprehensiveness and validity of the extracted
information pieces \cite{McCullough_etal_2007}. This step was used
to inform the synthesis and determined to what extent the synthesis
should rely on an extract. Given the interdisciplinary nature of medical
ethics research, we took into consideration that arguments in this
field often do not only rely on purely normative considerations but
include so called mixed judgements which encompass factual or empirical
claims as well as normative statements to arrive at a conclusion \cite{Schleidgen_etal_2022}.
We hence modified the approach of McCullough to reflect this based
on current proposals of reporting empirically informed ethical judgements.
For the final step we followed McDougall who suggests to understand
quality appraisal as integral part of the reporting of normative systematic
reviews \cite{McDougall_2015}.  We hypothesized that this stage should
rely on content-based criteria of critical thinking and to report
strength and limitations of claims made in our dataset as well as
to contextualize findings with the wider literature if available.
Table~\ref{tab:assessment} shows the leading questions of our assessment in each stage.
The first two steps were carried out by the first author {[}J.H.{]}
and were supervised by {[}S.P-H{]}. The third step was carried out
in a joint review session with all authors.

\section*{Results}

\subsection*{Description of included studies}

The initial search resulted in 1,723 records with 955 remaining after
removal of duplicates; 795 records were excluded during title and
abstract screening.\begin{figure*}[!tbh]
\caption{\label{fig1:prisma}Flow of studies through the screening process}
\includegraphics[width=\textwidth]{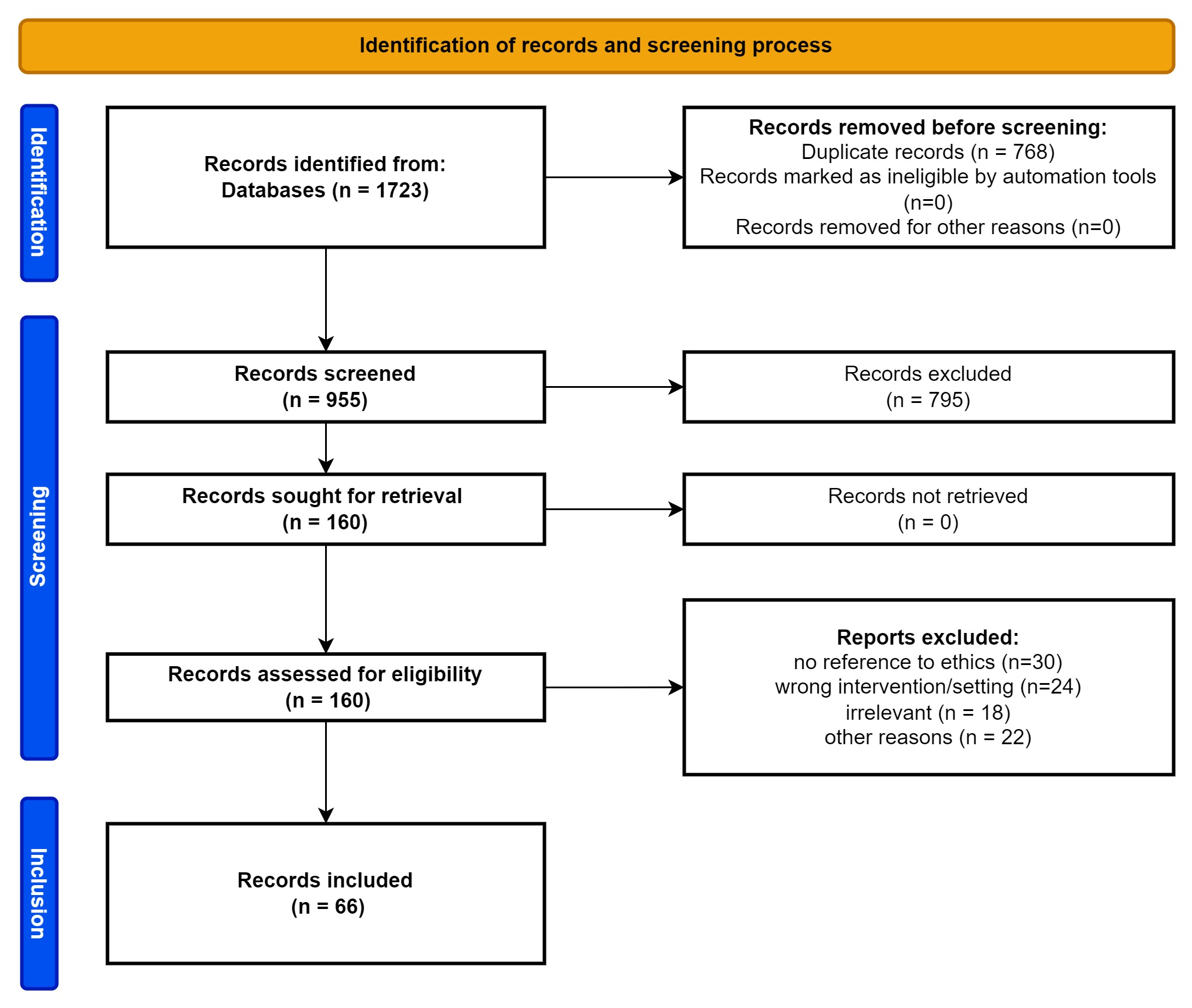}
\end{figure*} For 160 records, the full text was accessed. 94 were excluded, mostly
because of missing reference to ethical issues and wrong settings
or interventions. 66 records were included in the final dataset. The
flow of records through the screening process can be seen in Figure~\ref{fig1:prisma}. Our dataset included 27 original articles, six book chapters ,
11 reviews, eight comments, editorials or letters, eight perspective
articles and six articles of unclear type. 45 publications were published
in the field of medicine including 24 in surgery.
Ten publications came from the field of ethics and applied ethics
and neighbouring fields. Eight works were published in the field of
the technical sciences. Two studies came from social and behavioural
sciences. One work was of other origin. Publication dates ranged from
2000 to 2022. Table~\ref{tab:records}  gives an overview. Table~\ref{tab:quality}  shows the procedural
quality criteria as outlined above.

\begin{figure*}[bp]
\includegraphics[width=\textwidth]{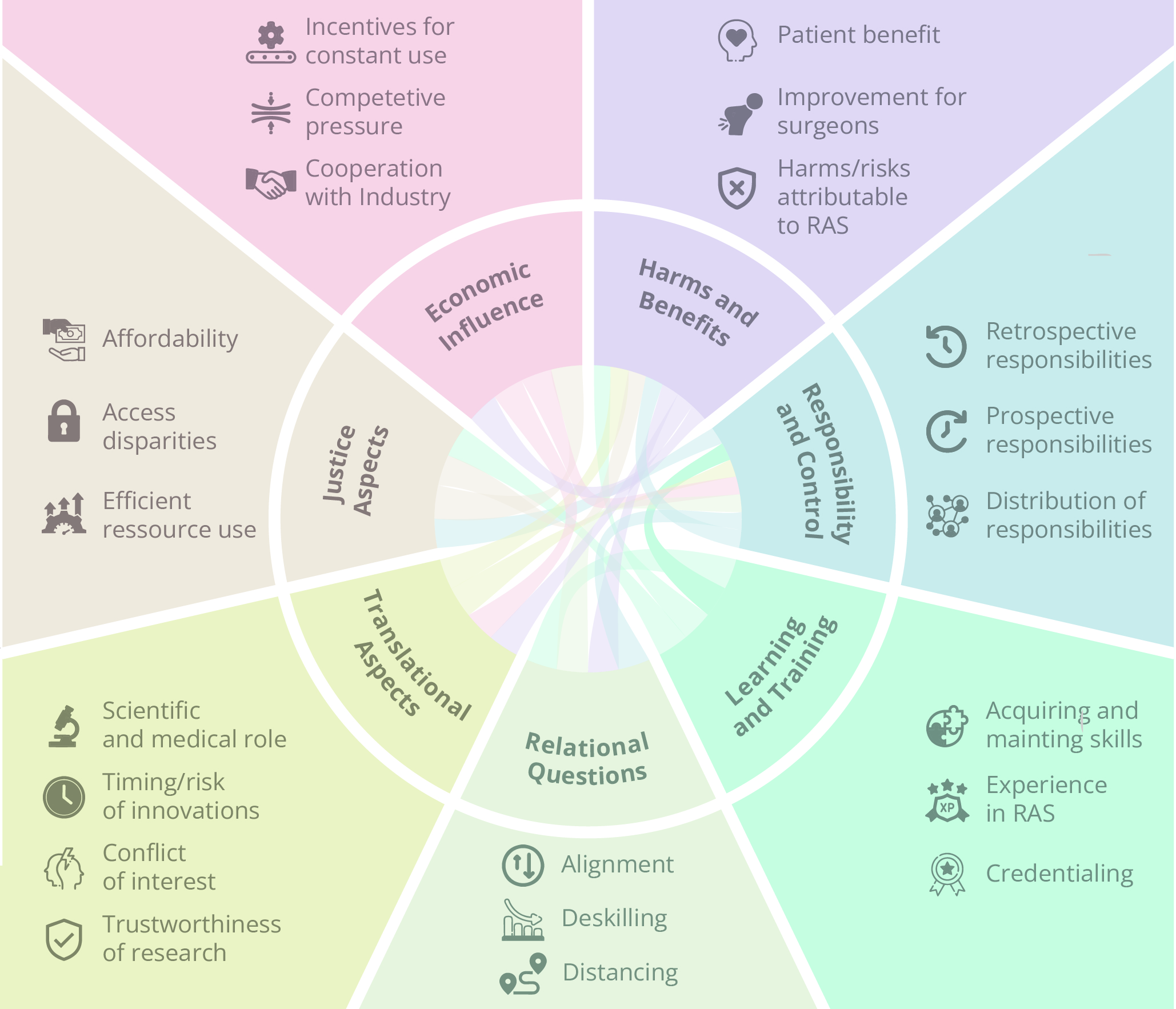}

\caption{\label{fig2:landscape}The Landscape of ethical issues in RAS}
\end{figure*}

With our analysis 7 major strands of the debate emerged. These include
questions of harms and benefits, responsibility and control, the professional-patient-relationship,
ethical questions of surgical training and learning, questions of
justice, translational questions and questions regarding economic
values. In what follows, we base our report on these major strands
noting a tight interconnectedness of all themes that is illustrated
in Figure~\ref{fig2:landscape}. We use these connections to further structure our report
and provide a consistent (though surely not the only possible) narrative.

\subsection*{Mapping the ethical landscape}

\subsubsection*{Benefits and harms to patients and surgeons}

With regard to patient benefit, it is agreed that it is an important
ethical criterion in support of the use of RAS \cite{Decker_2018b,Datteri_2013b}.
Wightman et al. argue that patient benefit should be the primary evaluation
criterion from an ethical perspective \cite{Wightman_etal_2020}.
The assumed benefits include, for example, lower rates of postoperative
complications \cite{Platis_Zoulias_2014}, a shortened length of stay
in hospital, a reduction of care expenses \cite{Esperto_etal_2022},
less invasive procedures \cite{Platis_Zoulias_2014,Angelos_2016}
as well as an overall increase in quality of surgical procedures through
standardization and prevention of outliers in performance \cite{Narsinh_etal_2022,Cavinatto_etal_2019}.
In addition, these devices may enable more complex procedures that
would be difficult or impossible to handle with traditional methods
\cite{Wightman_etal_2020,Giovagnoli_etal_2019,Nestor_Wilson_2019b}
and may increase precision of surgical maneuvers \cite{Saniotis_Henneberg_2021}.
In some cases, further benefits may extend beyond the procedure itself,
such as a decreased exposure to radiation or a reduced risk of infections
\cite{Platis_Zoulias_2014,Narsinh_etal_2022}. However, Decker cautions
that evaluations of these benefits must be highly specific to each
surgical domain and should not be generalized \cite{Decker_2018b}.
This caution is echoed by several authors noting an unclear situation
regarding whether these assumed benefits can actually be supported
by evidence in specific cases 
\cite{Sharkey_Sharkey_2012,Teller_etal_2022,Smyth_etal_2013}.
Hutchinson et al. note a considerable debate with regard to these
issues, concluding that establishing superiority of robot-assisted
procedures beyond doubt may be a challenge \cite{Hutchison_etal_2016b}.
According to Scheetz and Dimick, commenting on a safety communication
released by the United States Food and Drug Administration (FDA) \cite{_2024},
there is little robust evidence suggesting that robot-assisted surgery
is superior to laparoscopic or open procedure \cite{Sheetz_Dimick_2019,Lirici_2022}.
While the FDA communication is mainly concerned with mastectomy and
other cancer-related surgery, the authors cite further evidence to
generalize this claim.
\singlespacing
\tablebodyfont%
\addtocounter{table}{-1}
\renewcommand\footnotetext[2][]{{\removelastskip\vskip3pt%
\let\tablebodyfont\tablefootnotefont%
\hskip0pt\if!##1!\else{\smash{$^{##1}$}}\fi##2\par}}%
\begin{tabularx}{\textwidth}{@{\extracolsep\fill}
L{\dimexpr.17\textwidth-2\tabcolsep}
L{\dimexpr.33\textwidth-2\tabcolsep}
L{\dimexpr.25\textwidth-2\tabcolsep}
L{\dimexpr.08\textwidth-2\tabcolsep}
L{\dimexpr.07\textwidth-2\tabcolsep}
L{\dimexpr.1\textwidth-2\tabcolsep}}
\tabularxcaption{\label{tab:records}Table~\ref*{tab:records}}{Overview on the included records.}\\
\toprule%
Title & Focus of study & Journal category\textsuperscript{*}  & Article type\textsuperscript{**} & Year & COI \\   
\midrule
\endfirsthead
\tabularxcaption{Table~\ref*{tab:records}}{Overview on the included records (continued).}\\
\toprule%
Title & Focus of study & Journal category\textsuperscript{*}  & Article type\textsuperscript{**} & Year & COI \\   
\midrule
\endhead
\\  \botrule
\multicolumn{6}{l}{\parbox{\dimexpr1\textwidth-2\tabcolsep}{ \textsuperscript{*} \footnotesize{By Web of Science Journal Classification (n.a. is self-assessed)}\newline \textsuperscript{**}  \footnotesize{P/V = Perspective/Viewpoint, E = Editorial, R = Review, Ch = Chapter, OA(e) = Original Article (empricial), OA(t) = Original Article (theoretical), L = Letter, C = Comment, O = Other}}}\\
\multicolumn{6}{r}{\footnotesize{continued on the next page}}\\ 
\multicolumn{6}{l}{}
\endfoot
\\  \botrule 
\multicolumn{6}{l}{\parbox{\dimexpr1\textwidth-2\tabcolsep}{ \textsuperscript{*} \footnotesize{By Web of Science Journal Classification (n.a. is self-assessed)}\newline \textsuperscript{**}  \footnotesize{P/V = Perspective/Viewpoint, E = Editorial, R = Review, Ch = Chapter, OA(e) = Original Article (empricial), OA(t) = Original Article (theoretical), L = Letter, C = Comment, O = Other}}}\\
\endlastfoot
Angelos \cite{AngelosP._2015} &Surgical ethics, RAS procedure &N.a. (Endocrinology, surgery) &P/V & 2015  &No\\
Angelos \cite{Angelos_2016} &Ethics &Surgery &P/V & 2016  &No\\
Angelos \cite{Angelos_2017} &Ethics &Surgery &E & 2017  &Unclear\\
Anvari \cite{Anvari_2005} &Telesurgery, surgical robots &Respiratory System; Surgery; Cardiac and Cardiovascular System &R & 2005  &Unclear\\
Bendel \cite{Bendel_2015} &Ethics, Medical Machine Ethics &N.a. (Medical Ethics) &Ch & 2015  &Unclear\\
Cavinatto et al. \cite{Cavinatto_etal_2019} &Conflict of Interest, RAS Procedure, Medical Ethics &Orthopedics &OA(e) & 2019  &Yes\\
Chitwood \cite{Chitwood_2019} &Ethics &Surgery; Cardiac and Cardiovascular System; Respiratory System &Co & 2019  &No\\
Collins et al. \cite{Collins_etal_2020} &Telesurgery, Education, Patient Safety &Urology and Nephrology &R & 2020  &Yes\\
Criss et al. \cite{Criss_etal_2019} &Conflict of Interest, Funding &Surgery &R & 2019  &Unclear\\
Datteri \cite{Datteri_2013b} &Medical Robotics, Robot Ethics, Philosophy of Science &Ethics &OA(t) & 2013  &Unclear\\
De Togni et al. \cite{Togni_etal_2021} &Artificial Intelligence, Health and Care, Robot Ethics &Public, Enivironmental and Occupational Health &OA(t) & 2021  &Unclear\\
Decker \cite{Decker_2018b} &Robotics, Ethics, Hospital &N.a. (Medical Ethics) &OA(t) & 2018  &Unclear\\
DeFrance et al. \cite{DeFrance_etal_2021} &RAS Procedure, Outcomes, &Orthopedics &R & 2021  &Yes\\ 
Di Paolo et al. \cite{DiPaolo_etal_2019} &RAS, inequality, Informed Consent &Surgery &O & 2019  &No\\
Dickens \& Cook \cite{Dickens_Cook_2006} &RAS, Legal issues, Ethics across borders &Obstetrics and Gynecology &OA(t) & 2006  &Unclear\\
El-Bahnasawi et al. \cite{ElBahnasawi_etal_2019} &Innovation, Training &Surgery; Gastroenterology and Hepatology &L & 2019  &No\\
Esperto et al. \cite{Esperto_etal_2022} &RAS in Urology, Ethical challenges &Urology and Nephrology &R & 2022  &No\\
Ficuciello et al. \cite{Ficuciello_etal_2019} &RAS, Ethics, Human Control &N.a. (Behavioural Robotics) &OA(t) & 2019  &Unclear\\
Geiger \& Hirschl \cite{Geiger_Hirschl_2015}  &Innovation, RAS, Ethics &Surgery; Pediatrics &OA(t) & 2015  &Unclear\\
Giovagnoli et al. \cite{Giovagnoli_etal_2019} &RAS, Robotics, Roboethics &Philosophy &OA(t) & 2019  &Unclear\\
Graur et al. \cite{Graur_etal_2010} &Ethics, RAS, telesurgery &N.a. (Computer Science, Hardware) &Ch & 2010  &Unclear\\
Hung et al. \cite{Hung_etal_2018} &RAS, telemedicine, education &Urology and Nephrology &R & 2018  &No\\
Hutchison et al. \cite{Hutchison_etal_2016b} &Justice, Innovation, Ethics &Medical Ethics &OA(t) & 2016  &Unclear\\
Jimbo et al. \cite{Jimbo_etal_2019} &Conflict of Interest, RAS in Urology, Robotics &Urology and Nephrology &OA(e) & 2019  &Yes\\
Jones \& McCullough \cite{Jones_McCullough_2002} &Ethics &Surgery; Peripheral Vascular Diseases &O & 2002  &Unclear\\
Lam et al. \cite{Lam_etal_2021b} &Digital Surgery, Robotics, Ethics &Health Care Sciences and Services; Medical Informatics &P/V & 2021  &Yes\\
Larson et al. \cite{LarsonJ.A._etal_2014} &Ethics, Patient Safety &Surgery &OA(t) & 2014  &Yes\\
Lee Char et al. \cite{LeeChar_etal_2013} &Informed Consent, Innovation &Surgery &OA(e) & 2013  &No\\
Lin et al. \cite{Lin_etal_2012} &Robotics &N.a. (Technology, engineering, machinery) &Ch & 2012  &Unclear\\
Lirici \cite{Lirici_2022} &RAS, Innovation &Surgery &E & 2022  &No\\
Mavroforou et al. \cite{Mavroforou_etal_2010} &Robotics, Surgery, Ethics &Peripheral Vascular Diseases &O & 2010  &Unclear\\
Narsinh et al. \cite{Narsinh_etal_2022} &Robotics, Ethics, Telemedicine &Neuroimaging &OA(t) & 2022  &Yes\\
Nestor \& Wilson \cite{Nestor_Wilson_2019b} &Robotics, Surgery, Anticipatory Ethics &Ethics &OA(t) & 2019  &Unclear\\
Patel et al. \cite{Patel_etal_2015b} &RAS, Spin &Gastroenterology and Hepatology; Surgery &OA(e) & 2015  &Yes\\
Patel et al. \cite{Patel_etal_2018} &RAS, Conflict of Interest &Surgery &OA(e) & 2018  &Yes\\
Platis, Zoulias \cite{Platis_Zoulias_2014} &Healthcare Services, RAS, Administration &N.a. (Social and behavioural sciences) &OA(t) & 2014  &Unclear\\
Polk et al. \cite{Polk_etal_2019} &Conflict of Interest, Guidelines &Gastroenterology and Hepatology; Surgery &P/V & 2019  &Yes\\
Saniotis \& Henneberg \cite{Saniotis_Henneberg_2021} &Neurosurgical Robot systems, Moral Agency, Bioethics &N.a. (Ethics) &OA(t) & 2021  &No\\
Satava \cite{Satava_2002} &Legal Issues, Ethical Issues &N.a. (Surgery) &OA(t) & 2002  &Unclear\\
Schlottmann \& Patti \cite{Schlottmann_Patti_2017} &Simulation, Training, RAS &Surgery &L & 2017  &No\\
See et al. \cite{SeeW.A._etal_2014} &RAS, Commercial Endorsement, Case Volumes &Oncology; Urology and Nephrology &OA(t) & 2014  &Unclear\\
Senapati \& Advincula \cite{SenapatiS._AdvinculaA.P._2005} &Robotics, Telesurgery, Gynecologic Surgery &Obstetrics and Gynecology &P/V & 2005  &Unclear\\
Shahzad et al. \cite{Shahzad_etal_2019} &Robotics, Tele-Robotics, Ethical Issues &Medicine, General and Internal &R & 2019  &No\\
Sharkey \& Sharkey \cite{Sharkey_Sharkey_2012}  &RAS, Ethics &N.a. (Ethics) &Ch & 2012  &No\\
Sharkey \& Sharkey \cite{Sharkey_Sharkey_2013}  &RAS, Ethical framework &Computer Science, Hardware and Architecture; Computer Science, Software Engineering &OA(t) & 2013  &No\\
Sheetz \& Dimick \cite{Sheetz_Dimick_2019} &RAS, Patient Safety, Learning &Medicine, General and Internal &P/V & 2019  &Unclear\\
Siciliano \& Tamburrini \cite{Siciliano_Tamburrini_2019} &Robotics, Ethics, Human-Machine-Interaction &Religion &OA(t) & 2019  &Unclear\\
Siqueira-Batista et al. \cite{SiqueiraBatistaR._etal_2016} &Bioethics, Surgery, Robotics &Gastroenterology and Hepatology &R & 2016  &No\\
Smith \cite{Smith_2013} &RAS, Responsibility &Urology and Nephrology &E & 2013  &Unclear\\
Smyth et al. \cite{Smyth_etal_2013} &RAS, Surgical Ethos &Surgery; Respiratory System; Cardiac and Cardiovascular System &P/V & 2013  &No\\
Spiers et al. \cite{Spiers_etal_2022} &RAS Kidney transplant, minimally invasive surgery &Urology and Nephrology &R & 2022  &Yes\\
Spillman \& Sade \cite{Spillman_Sade_2014b} &Ethics &N.a. (Medical Ethics) &O & 2014  &Unclear\\
Stanberry \cite{Stanberry_2000b} &Telesurgery and Tobotics, health telematics &Medicine, General and Internal &OA(t) & 2000  &Unclear\\
Steil et al. \cite{Steil_etal_2019} &Robots, team-machine-interaction, hybrid action &Health Care Sciences and Services; Medical Informatics; Computer Science, Information Systems &O & 2019  &No\\
Strong et al. \cite{Strong_etal_2014} &Surgical Education, Human/Robotic, Costs &Surgery &O & 2014  &Yes\\
Sullins \cite{Sullins_2014b} &Ethics, Trust, RAS &N.a. (Philosophy) &OA(t) & 2014  &Unclear\\
Swarnalatha \& Menon \cite{Swarnalatha_Menon_2022} &Telerobotics, medical robotics &N.a. (Engineering, information technology) &OA(t) & 2022  &Unclear\\
Teller et al. \cite{Teller_etal_2022} &RAS, Innovation, Ethics &Surgery; Oncology &OA(t) & 2022  &No\\
Thomas et al. \cite{Thomas_etal_2020} &Robotics, Advertising, Surgery &Surgery &R & 2020  &Yes\\
Tzafestas \cite{Tzafestas_2016} &Robotics &N.a. (Engineering, information technology) &Ch & 2016  &Unclear\\
van der Waa et al. \cite{vanderWaa_etal_2020} &Moral Decision Making, Human-Agent Teaming, Machine Ethics &N.a. (Engineering, behavioural sciences) &Ch & 2020  &Unclear\\
Vilanilam \& Venkat \cite{Vilanilam_Venkat_2022} &Robotic Neurosurger, Ethics, Legal Issues &Surgery; Clinical Neurology &E & 2022  &No\\
Whiteside \cite{Whiteside_2008} &Robotic Gynecologic Surgery &Obstetrics and Gynecology &E & 2008  &Yes\\
Wightman et al. \cite{Whiteside_2008} &Ethics, RAS, informed consent &Surgery &R & 2020  &Yes\\
Woo et al. \cite{Woo_etal_2019} &Informed Consent, Surgeon Experience &Surgery; Cardiac and Cardiovascular System; Respiratory System &P/V & 2019  &No\\
Zorn et al. \cite{Zorn_etal_2009} &Urologic Robotic Surgery, Training, Credentialing &Urology and Nephrology &OA(t) & 2009  &Yes\\
\end{tabularx}
\normalsize

Authors disagree on the reasons for the perceived lack of empirical
evidence supporting arguments based on patient benefit. In many cases,
determining specific outcomes (e.g. beyond decreased morbidity and
mortality) is challenging \cite{AngelosP._2015}. Whether, for example,
advantages such as improved accuracy of maneuvers correlate with better
patient outcomes is uncertain \cite{DeFrance_etal_2021}. Furthermore,
a lack of scientific quality of existing studies is noted. According
to Scheetz and Dimick, for example, most studies in the field have
only been small, single-center trials without effective controls \cite{Cavinatto_etal_2019,Sheetz_Dimick_2019},
while long term data is often not available. Consequently, there is
little evidence with regard to patient benefits \cite{Sheetz_Dimick_2019,Lirici_2022,Vilanilam_Venkat_2022}.
Another noted factor concerns the trustworthiness of research about
RAS \cite{Tzafestas_2016}, which may be influenced by the close intertwinement
of researchers and industry. Failure to disclose financial connections
and conflict of interest is also common and contributes to a lack
of transparency \cite{Criss_etal_2019,Jimbo_etal_2019,Polk_etal_2019,Patel_etal_2015b,PatelR._etal_2020}.
Several authors report correlations between ties to the industry and
more favourable study outcomes or a tendency to overinterpret results
and exaggerate claims \cite{Criss_etal_2019,PatelR._etal_2020,Patel_etal_2015b}.
Although the validity and interpretation of these findings is itself
subject to debate \cite{Tanaka_2019,Elizondo_Koh_2019}, it shows
ongoing concerns with regard to transparency and trustworthiness of
empirical results. 

The potential of harm for patients is primarily discussed in terms
of risks to physical integrity \cite{Datteri_2013b,Wightman_etal_2020,Esperto_etal_2022,Sharkey_Sharkey_2012,Sheetz_Dimick_2019,Vilanilam_Venkat_2022,Lin_etal_2012,Geiger_Hirschl_2015}.
Although potential harm is a frequently cited, the exact nature of
these risks remain unclear and are only roughly indicated. Wightman
et al. argue that devices should not add additional harm (or the risk
of it) to a procedure \cite{Wightman_etal_2020}. Discussing a specific
procedure, Angelos delivers an example, noting that robotic procedures
can be connected to very specific risks that come with the positioning
of patients and the point of access of the instruments that is unique
to RAS. As Angelos concludes, this results in specific risks that
are directly attributable to the use of the device \cite{AngelosP._2015}.
Sharkey notes increased morbidity and mortality as well as longer
hospital stays but does not provide backing evidence for these claims
\cite{Sharkey_Sharkey_2013}. Sheetz and Dimick cite evidence for
radical hysterectomy showing a significant decrease in long-term survival
rates \cite{Sheetz_Dimick_2019}. Others note that harm due to malfunctions
can only be observed in a minority of cases \cite{Datteri_2013b,Esperto_etal_2022}.
Di Paolo et al. nevertheless caution that 77\% percent of adverse
events in RAS with DaVinci Systems could be rated as device-specific
incidents, that is, situations in which the device played a significant
and causally relevant role \cite{DiPaolo_etal_2019}. This claim is
based on data from 2009 to 2012, including all adverse events reported
in the FDA MAUDE Database. However, as this data also includes incidents
occurring in preparation of surgery, we
find that the cited evidence does not yield direct insights into the
question as to how many incidents in which harm was very likely can
be attributed to the devices. We conclude that it does not support Di Paolo et al’s
claim. Notwithstanding this, the data shows that at least around 25\%
of all incidents have moderate (17,73\%), severe (4,44\%) or life-threatening
outcomes (2,89\%) implying patient involvement. It is, hence, still
likely that in a considerable amount of negative outcomes a causal
connection to the use of RAS exists \cite{Gupta_etal_2017}. In addition,
Geiger and Hirschl suggest, based on empirical evidence, that complications
and malfunctions are frequently underreported \cite{Geiger_Hirschl_2015}. 

Beyond physical risks, informational harms are also widely discussed
in the literature. These include threats to privacy, confidentiality
or security breaches are primarily connected to data collection and
use that comes with the devices \cite{Sharkey_Sharkey_2013,Shahzad_etal_2019,Spillman_Sade_2014b}.
Data security is mostly discussed within the context of telesurgery, a subspeciality 
of RAS, in which the distance between the control
console and the actuator extends the usual distances and can even
expand beyond national borders \cite{Sharkey_Sharkey_2012,Sharkey_Sharkey_2013,Anvari_2005}.
It should be noted, however, that the underlying cause for security
vulnerabilities is the transmission of data between device and control.
Remote distance may therefore add only additional layers of vulnerability
and complexity to an already existing problem \cite{Gordon_etal_2022}.
In this regard, Vilaniam and Venkat argue that clinical data security
concerns arise whenever third parties are involved that could access,
exploit or misuse data \cite{Vilanilam_Venkat_2022}.
\singlespacing
\tablebodyfont%
\renewcommand\footnotetext[2][]{{\removelastskip\vskip3pt%
\let\tablebodyfont\tablefootnotefont%
\hskip0pt\if!##1!\else{\smash{$^{##1}$}}\fi##2\par}}%
\begin{tabularx}{\textwidth}{@{\extracolsep\fill}
L{\dimexpr.4\textwidth-1\tabcolsep}
L{\dimexpr.2\textwidth-1\tabcolsep}
L{\dimexpr.2\textwidth-1\tabcolsep}
L{\dimexpr.2\textwidth-1\tabcolsep}}
\tabularxcaption{\label{tab:quality}Table~\ref*{tab:quality}}{Overview on procedural quality control criteria.}\\
\toprule%
Paper &Peer reviewed &Minor work\textsuperscript{*} &Material richness \\  
\midrule
\endfirsthead
\tabularxcaption{Table~\ref*{tab:quality}}{Overview on procedural quality control criteria (continued).}\\
\toprule%
Paper &Peer reviewed &Minor work\textsuperscript{*} &Material richness \\ 
\midrule
\endhead
\\  \botrule 
\multicolumn{4}{l}{\parbox{\dimexpr1\textwidth-2\tabcolsep}{ \textsuperscript{*} \footnotesize{Minor work includes comments, letters and editorials.}}}\\
\multicolumn{4}{r}{\footnotesize{continued on the next page}}
\endfoot
\\  \botrule 
\multicolumn{4}{l}{\parbox{\dimexpr1\textwidth-2\tabcolsep}{ \textsuperscript{*} \footnotesize{Minor work includes comments, letters and editorials.}}}\\
\endlastfoot
Angelos \cite{AngelosP._2015} &Yes &No &Yes\\
Angelos \cite{Angelos_2016} &Yes &No &Yes\\
Angelos \cite{Angelos_2017} &No &Yes &Yes\\
Anvari \cite{Anvari_2005} &Yes &No &Yes\\
Bendel \cite{Bendel_2015} &Unclear &No &No\\
Cavinatto et al. \cite{Cavinatto_etal_2019} &Yes &No &Yes\\
Chitwood \cite{Chitwood_2019} &No &Yes &No\\
Collins et al. \cite{Collins_etal_2020} &Yes &No &No\\
Criss et al. \cite{Criss_etal_2019} &Yes &No &Yes\\
Datteri \cite{Datteri_2013b} &Yes &No &Yes\\
De Togni et al. \cite{Togni_etal_2021} &Yes &No &Yes\\
Decker \cite{Decker_2018b} &Yes &No &Yes\\
DeFrance et al. \cite{DeFrance_etal_2021} &Yes &No &Yes\\
Di Paolo et al. \cite{DiPaolo_etal_2019} &No &Yes &Yes\\
Dickens \& Cook \cite{Dickens_Cook_2006} &Yes &No &No\\
El-Bahnasawi et al. \cite{ElBahnasawi_etal_2019} &No &Yes &No\\
Esperto et al. \cite{Esperto_etal_2022} &Yes &No &Yes\\
Ficuciello et al. \cite{Ficuciello_etal_2019} &Unclear &No &Yes\\
Geiger, Hirschl \cite{Geiger_Hirschl_2015} &Yes &No &Yes\\
Giovagnoli et al. \cite{Giovagnoli_etal_2019} &Unclear &No &No\\
Graur et al. \cite{Graur_etal_2010} &Unclear &No &Yes\\
Hung et al. \cite{Hung_etal_2018} &Yes &No &No\\
Hutchison et al. \cite{Hutchison_etal_2016b} &Yes &No &Yes\\
Jimbo et al. \cite{Jimbo_etal_2019} &Yes &No &No\\
Jones \& McCullough \cite{Jones_McCullough_2002} &No &Yes &No\\
Lam et al. \cite{Lam_etal_2021b} &No &Yes &No\\
Larson et al. \cite{LarsonJ.A._etal_2014} &Yes &No &No\\
Lee Char et al. \cite{LeeChar_etal_2013} &Yes &No &Yes\\
Lin et al. \cite{Lin_etal_2012} &Unclear &No &No\\
Lirici \cite{Lirici_2022} &No &Yes &No\\
Mavroforou et al. \cite{Mavroforou_etal_2010} &No &Yes &No\\
Narsinh et al. \cite{Narsinh_etal_2022} &Yes &No &Yes\\
Nestor \& Wilson \cite{Nestor_Wilson_2019b} &Yes &No &Yes\\
Patel et al. \cite{Patel_etal_2015b} &Yes &No &No\\
Patel et al. \cite{Patel_etal_2018} &Yes &No &No\\
Platis \& Zoulias \cite{Platis_Zoulias_2014} &Unclear &Yes &No\\
Polk et al. \cite{Polk_etal_2019} &Yes &Yes &Yes\\
Saniotis \& Henneberg \cite{Saniotis_Henneberg_2021} &Yes &No &Yes\\
Satava \cite{Satava_2002} &Unclear &No &No\\
Schlottmann \& Patti \cite{Schlottmann_Patti_2017} &No &Yes &Yes\\
See et al. \cite{SeeW.A._etal_2014} &Yes &No &No\\
Senapati \& Advincula \cite{SenapatiS._AdvinculaA.P._2005} &No &Yes &No\\
Shahzad et al. \cite{Shahzad_etal_2019} &Unclear &No &No\\
Sharkey \& Sharkey \cite{Sharkey_Sharkey_2012} &No &No &Yes\\
Sharkey \& Sharkey \cite{Sharkey_Sharkey_2013} &Yes &No &Yes\\
Sheetz \& Dimick \cite{Sheetz_Dimick_2019} &Yes &Yes &No\\
Siciliano \& Tamburrini \cite{Siciliano_Tamburrini_2019} &Yes &No &Yes\\
Siqueira-Batista et al. \cite{SiqueiraBatistaR._etal_2016} &Yes &No &No\\
Smith \cite{Smith_2013} &No &Yes &No\\
Smyth et al. \cite{Smyth_etal_2013} &Yes &No &No\\
Spiers et al. \cite{Spiers_etal_2022} &Yes &No &No\\
Spillman \& Sade \cite{Spillman_Sade_2014b} &Unclear &No &No\\
Stanberry \cite{Stanberry_2000b} &No &No &No\\
Steil et al. \cite{Steil_etal_2019} &Yes &No &Yes\\
Strong et al. \cite{Strong_etal_2014} &No &No &Yes\\
Sullins \cite{Sullins_2014b} &Unclear &No &Yes\\
Swarnalatha \& Menon \cite{Swarnalatha_Menon_2022} &Yes &No &No\\
Teller et al. \cite{Teller_etal_2022} &Yes &No &Yes\\
Thomas et al. \cite{Thomas_etal_2020} &Yes &No &No\\
Tzafestas \cite{Tzafestas_2016} &Unclear &No &No\\
van der Waa et al. \cite{vanderWaa_etal_2020} &Unclear &No &Yes\\
Vilanilam \& Venkat \cite{Vilanilam_Venkat_2022} &No &Yes &No\\
Whiteside \cite{Whiteside_2008} &No &Yes &No\\
Wightman et al. \cite{Whiteside_2008} &Yes &No &Yes\\
Woo et al. \cite{Woo_etal_2019} &No &Yes &No\\
Zorn et al. \cite{Zorn_etal_2009} &No &No &No\\

\end{tabularx}
\normalsize
Beyond patient benefit and harm, the assumed benefits for surgeons
are also considered part of the ethical calculus \cite{Narsinh_etal_2022,Giovagnoli_etal_2019,Teller_etal_2022,Steil_etal_2019,Angelos_2017}.
This includes, for example, improvement of physician skills (e.g.
suppression of hand tremor, less radiation exposure), avoidance of
errors through fatigue, inattentiveness, and stress as well as shortened
learning curves in some procedures. Based on a case example, Angelos
\cite{Angelos_2017} argues, that in some cases, empirical data supports
concluding that using a robotic procedure may benefit only the surgeon,
for example, as a way to train and maintain their skills or to gain
experience with a less demanding procedure, while other manual options
with equal outcomes would be available to the patient. This raises
the additional question whether it would be acceptable to offer such
a procedure to a patient purely for the benefit of the surgeon \cite{Angelos_2017}.

\subsubsection*{Responsibility and control}

With regard to responsibility, most authors agree that being able
to assign responsibility is a normative requirement, no matter the
circumstances. Failing to be able to do so would otherwise result
in so-called responsibility gaps \cite{vanderWaa_etal_2020,Siciliano_Tamburrini_2019}.
This term refers to a situation in which no addressee of a moral claim
can be determined. Cases of responsibility gaps present serious ethical
problems. Some authors, however, consider it a challenge that the
detailed ethical interpretation of the responsibility principle, “raises
special ethical issues in the context of increasing autonomy of medical
robots where physicians are no longer in control of each and every
aspect of medical procedures on the human body” \cite{Siciliano_Tamburrini_2019}.

The primary variable in questions of responsibility is the ability
to exert any form of normatively meaningful control over the devices
and, thus, is tightly connected to their level of autonomy \cite{Decker_2018b,Siciliano_Tamburrini_2019}.
In this context, autonomy does not denote a moral good worthy of protection
but is understood in a descriptive sense as a functional capacity.
Nyholm defines functional autonomy as the degree to which a device
is capable of creating a representation of its surroundings and to
make decisions how to reach a predefined goal \cite{Nyholm_2020}.
With increasing degrees of functional autonomy, that is, increasing
active potential, questions of responsibility become more pressing
\cite{Saniotis_Henneberg_2021,Steil_etal_2019,vanderWaa_etal_2020,Siciliano_Tamburrini_2019}.

Various typologies and categories of robotic surgical devices have
been proposed by authors of our dataset to highlight significant differences
in these capacities. These include categories ranging from passive
to active in which robots are increasingly involved in more invasive
tasks \cite{Togni_etal_2021}, different modes of control (planning
and execution vs. supervised mode) \cite{Narsinh_etal_2022,Saniotis_Henneberg_2021,Ficuciello_etal_2019}
or different forms of supervision \cite{Sharkey_Sharkey_2012,Nestor_Wilson_2019b}.
This range is also mirrored in respective conceptual language in addressing
the devices, from “slave systems” or “master-slave systems” to the
ascription of agent-like properties (e.g. “new players”, “mediators”)
\cite{Wightman_etal_2020,Narsinh_etal_2022,Anvari_2005,Siciliano_Tamburrini_2019,Ficuciello_etal_2019}.
Others refer to the classification of Yang et al. noted above \cite{Siciliano_Tamburrini_2019}.
All these classifications aim to highlight morally relevant differences
in the ability to exercise control over the devices and set the stage
for in-depth inspections of environmental conditions or human-machine
requirements for control, as well as specific obligations under specified
circumstances.

A number of authors in this strand of the debate devote their work
to exploring the topic of retrospective responsibility \cite{Siciliano_Tamburrini_2019,Togni_etal_2021,Ficuciello_etal_2019,Datteri_2013b,Decker_2018b}.
Retrospective responsibility is concerned with an ascription of responsibilities
after the facts, that is, when harm to the patient has occurred during
a procedure or a step of procedure. Datteri argues that, in this case,
someone needs to be identified to be blamed (i.e. bearing, what Datteri
understands as, moral responsibility), and someone (else) should be
held liable for compensating the damage \cite{Datteri_2013b}. The
argument seems to imply a distribution or compartmentalization of
different functions of responsibility. Others add that duties to compensate
need to be distributed evenly and fairly between different parties
involved, such as hospitals, manufacturers or human agents \cite{Siciliano_Tamburrini_2019,Decker_2018b}.
However, it remains unclear what constitutes a fair distribution of
responsibilities and whether this introduces a normative criterion
to compensate for ascription problems and to avoid responsibility
gaps due to a lack of control.

On the prospective side of responsibility, that is, with regard to
the duties surgeons bear in preparation, implementation, and prospective
decision making, there are various perspectives. According to Wightman
and others, a surgeon’s prospective responsibility in participating
in RAS can be primarily determined on grounds of their duty to avoid
harm to the patient \cite{Wightman_etal_2020,Spillman_Sade_2014b}.
This includes the decision to offer a robotic surgical procedure
to a patient \cite{Angelos_2016}, setting appropriate conditions
and procedures for informed consent, being aware of the specific problems
connected to a robotic surgical approach, and possessing the ability
and readiness to step in and take full control in case of malfunctions
and errors \cite{Esperto_etal_2022}. 

\subsubsection*{Surgical learning and training}

Given the complexities of surgical learning \cite{Pakkasjarvi_etal_2024},
credentialing and standards for introducing new procedures, this field
includes arguments highlighting the responsibility of surgeons to
acquire new skills and contribute to the introduction of new procedures
\cite{Geiger_Hirschl_2015,Collins_etal_2022,LarsonJ.A._etal_2014,Zorn_etal_2009}.
The arguments heavily rely on the concept of a surgical learning curve
(LC) through case-based exposure and gradual increase in volume. LC
is defined as the span of time (the number of cases) required for
a surgeon to acquire competence in a procedure \cite{Esperto_etal_2022}.
Competence is defined as the proficiency to perform a new procedure
in reasonable time and with favorable outcomes, as well as by having
the expertise to react to unforeseen and unplanned events \cite{Esperto_etal_2022}. 

It is commonly agreed that the RAS-LC differs from other surgical
procedures. Transition from manual surgical procedures to RAS, requires
extensive knowledge and skills with regard to the former. However,
it also, typically requires experience around 5-20 cases with RAS
to become proficient, while the LC in total may extend up to 100 –
250 cases, depending on procedure, related experience of the surgeon,
training, attitude of the surgeon, and confidence \cite{Esperto_etal_2022,Hutchison_etal_2016b,Collins_etal_2022,LarsonJ.A._etal_2014,DiPaolo_etal_2019,Jones_McCullough_2002}.
Dimick and Sheetz argue that surgical credentialing programs typically
require a number of proctoring cases as proxy measure for competence,
which is considerably lower than what is needed to reach proficiency.
The authors report two cases to be sufficient in some RAS training
programs \cite{Sheetz_Dimick_2019}. This is in line with recent research
analyzing 42 training policies from the US and reporting a mean of
3.24 cases (with a range of 1-10 cases) for initial credentialing
\cite{Huffman_etal_2021}. Thus, a gap exists between acquiring credentials
to use RAS and acquiring true competency and expertise \cite{Sheetz_Dimick_2019,LarsonJ.A._etal_2014}. 

Secondly, irrespective of credentialing, procedures at an early stage
of the learning curve are inevitably connected to a higher risk compared
to later stages and may not meet the same quality standards \cite{Jones_McCullough_2002,LeeChar_etal_2013}.
In this way, surgical learning is closely intertwined with bioethical
questions of patient harm and benefit \cite{Esperto_etal_2022,Sharkey_Sharkey_2012,Geiger_Hirschl_2015,Sharkey_Sharkey_2013,Spillman_Sade_2014b,LarsonJ.A._etal_2014,Whiteside_2008}.
Weaknesses in standardized learning and credentialing procedures put
responsibility on surgeons to actively seek and acquire the necessary
skills and to ultimately decide when they are sufficiently prepared
to conduct procedures independently \cite{Smith_2013,Spiers_etal_2022,Angelos_2016}.

The two above mentioned categories of surgeons’ responsibilities and
considerations regarding the learning curve intersect especially when
it comes to procedures of informed consent. A key question which is
heavily debated is whether it is the surgeons’ responsibility to (proactively)
disclose their level of experience in procedures of informed consent. Most authors 
agree that surgeons should disclose their experience
with RAS to patients 
\cite{Wightman_etal_2020,Teller_etal_2022,Sheetz_Dimick_2019,AngelosP._2015,DiPaolo_etal_2019,LeeChar_etal_2013,Spiers_etal_2022,Sullins_2014b,Woo_etal_2019}.
 They conclude that the question as to whether surgeons feel sufficiently
trained and competent should be openly discussed alongside other important
information \cite{Spiers_etal_2022}. Its discussion with patients
is based in the duty to respect patient autonomy \cite{Woo_etal_2019}.
This line of argument is backed up by empirical data gathered by Lee
Char et al., who found that patients perceived it as essential to
know if the surgeon performs a procedure for the first time \cite{LeeChar_etal_2013}.
It should be noted, however, that this line of argument might raise
objections of double standards. It is a commonplace in medical practices
that an outcome varies with experience, while extensive self-reporting
of one’s experience is uncommon (and fallible). An argument would
therefore be needed to justify a deviation from these norms. In addition,
it remains, questionable whether a duty to disclose experience should
also apply to transitions from less demanding to more complex procedures
(or vice versa), which is common for the surgical process of learning
\cite{Sheetz_Dimick_2019}. Finally, Sullins notes that less experience
with RAS might also affect surgeons’ ability to effectively disclose
all relevant risks, as they might not be fully of aware of them \cite{Sullins_2014b}
rendering it questionable whether and to what extend such information
would add to the capacity of patients to decide autonomously.

\subsubsection*{Translational questions: Introducing new procedures}

The theme of translational questions is concerned with determining
ethically acceptable conditions for the introduction of new devices.
As noted above, the process of surgical innovation “falls into a middle
range between minor modifications and surgical research” \cite{Angelos_2016}.
It is described to occupy a grey zone with fewer regulations as compared
to other branches of medicine \cite{Vilanilam_Venkat_2022,Geiger_Hirschl_2015,Sullins_2014b}.
Within this grey zone, unethical experimentation – adopting new procedures
with the aim of contributing to a generalizable knowledge without
being able to reliably predict outcomes – must be avoided from an
ethical perspective \cite{Angelos_2016,Jones_McCullough_2002}. Responsibility
resides with the surgeon to adopt new practices to the benefit of
patients while at the same time refrain from using potentially dangerous
or harmful practices \cite{Sharkey_Sharkey_2012}. In addition, there
is a scientific responsibility to track outcomes, ensure appropriate
documentation and share this information. This obligation is often
seen as a first step in more systemic efforts to ease the pathway
of introducing a robotic surgical devices, where manufacturers, hospitals
and other institutions also carry additional responsibilities. It
implies carefully timing the surgical innovations: waiting until sufficient
data is available to support their use while not delaying its introduction
unnecessarily and risking depriving patients of potential benefits
\cite{Esperto_etal_2022,Teller_etal_2022,Vilanilam_Venkat_2022,Strong_etal_2014}.

Although the requirement of a reasonable balance of safety, benefit,
and timely adoption seems straightforward, determining all relevant
factors with sufficient certainty is challenging. Besides noted methodological
problems and data availability, the so-called Collingridge dilemma
adds further complication \cite{Steil_etal_2019}. In short, the Collingridge
dilemma highlights the observation that shaping technology is easier
in its early stages of adoption, but once established, it can become
almost impossible. On the other hand, however, in early phases uncertainty
about outcomes is typically very high and, hence, hampers potential
decision making processes, while corridors to influence technologies
in later stages with lesser uncertainties shrink or become even non-existent
\cite{Steil_etal_2019}.

\subsubsection*{Financial incentives and economic pressure}

As many authors note, the process of surgical innovation is not only
multifaceted and rich in itself but is also put under external pressure
emerging at the intersection of healthcare, and technology as business.
According to Sullins it can be argued that there should be a separation
of industry concerns from medical care. However, it also needs to
be acknowledged that this is hard to maintain in surgical practice
\cite{Sullins_2014b}. The fact that robotic surgical devices come
with a high monetary value attached and are introduced with the necessity
of generating profit translates into an ethically problematic influence
on the process of surgical innovation. Some authors express concerns
that in light of the high costs of devices’ maintenance and training,
adopting institutions have a strong incentive to further a constant
and broad use of RAS and may incentivize or even pressurize their
employees to comply to ensure adequate monetary revenue \cite{Angelos_2016,Sullins_2014b}.
Especially in market-oriented and competitive healthcare systems or
in competitive disciplines of surgery, further pressure may arise
to be able to compete with other entities or colleagues \cite{Saniotis_Henneberg_2021,Sharkey_Sharkey_2012,Sharkey_Sharkey_2013}.
Whether and to what extent such pressures actually exist is an empirical
question that is not addressed sufficiently in our dataset. However,
many authors argue that it must be part of surgeons to be aware of
and to withstand these pressures \cite{Angelos_2016}, thereby maintaining
an intact and trustworthy relationship with patients that is not unduly
influenced by economic motives.

While this argument puts surgeons in the position of gatekeepers to
the best of their patients as intrinsic part of their professional
responsibilities, many authors critically remark financial and cooperative
ties of surgeons with the industry. In addition, it has been noted
that aggressive direct-to-consumer advertising campaigns \cite{Thomas_etal_2020},
increased power of manufacturers through quasi-monopolization \cite{Esperto_etal_2022},
and misleading or exaggerated quality claims \cite{Sharkey_Sharkey_2012}
may pose ethical problems that may distort the conditions allowing
for decisions in the patients’ best interest.

\subsubsection*{Justice in RAS}

The factor of economic costs of RAS in connection with uncertain perspectives
on harm and benefit, raises complex questions from a perspective of
justice. The first question is, how to rate the fact that only some
medical facilities may be able to afford the investment in RAS devices.
Some authors point to empirical evidence showing that RAS interventions
can lead to inequalities and that patients are predominantly caucasian
male, located near teaching or university hospitals \cite{Esperto_etal_2022,Geiger_Hirschl_2015}.
However, as Decker notes, only if an analysis of costs and benefits
comes to a positive conclusion, it may be a requirement of justice
to make such an intervention accessible and affordable to all \cite{Decker_2018b}.
De Angelos argues more detailed that it would not be unjust to allow
RAS as a procedure of elective choice, meaning that patients would
have to bear additional costs, while it may constitute an injustice
if one was to believe that a RAS intervention would come with additial
benefits to patients \cite{AngelosP._2015}. 

Sharkey, again pointing to high costs of RAS, on the other hand argues
that a fair distribution of health care resources requires to limit
expenditures to interventions with proven safety and efficacy. Costly
surgical interventions like RAS may divert resources from more cost-effective
options. Teller et al. expand on this by suggesting that even if a
procedure proves effective and safe, it is important to consider who
will have to bear the additional costs to minimize inequities of care
or exclusion of patient populations \cite{Teller_etal_2022}.

Remarks like these set the stage for broader debates about distributive
justice as a matter of sharing benefits and burdens within a population
but also for a global justice perspective. Some authors find that
especially telesurgery programs could provide the opportunity to reach
distant locales and may alleviate many of the existing disparities
in the delivery of surgical care \cite{Anvari_2005,Graur_etal_2010}.
Hutchison et al, however, argue that RAS requires high investments
and constant training of surgeons to be cost-effective, which may
only be possible in densely populated areas \cite{Hutchison_etal_2016b}.
Finally, Anvari sees RAS and its telesurgical subfields as an opportunity
to connect to distant colleagues and to provide knowledge and support
in a longer run \cite{Anvari_2005}. Sullins, however objects by highlighting
that especially countries of the global south may not be in a position
to afford RAS and rely on more traditional approaches, making collegial
support and skill transfer more difficult \cite{Sullins_2014b}.

\subsubsection*{Relational aspects: human-machine and 
patient-professional 
relations}

Steil et al. and deTogni note that especially increased functional
autonomy may not only magnify questions of responsibility but may
also raise ethical questions of human-machine-interaction in a broader
sense \cite{Steil_etal_2019,Togni_etal_2021}. This includes considerations
around ethical team work, the human dimension in relations between
surgeon and patients, and the relation between health professionals
and other actors.

Beyond specific devices, ethical issues of alignment to technology
and growing dependencies are noted as potential ethical concerns.
Complex technical arrangements such as RAS are very likely to lead
to an alignment of surgical procedures with the requirements of robotic
systems, in which surgeons and team members need to acquire new skills
\cite{Steil_etal_2019,Togni_etal_2021}. This can lead to a form of
codependence – and on the negative side - may create strong incentives
to rely heavily on the machines’ capabilities. An example would be
a shift in burden of proof, when a surgeon wants to depart from a
suggested course of action by the machine and aims to overwrite partial
machine control \cite{Sullins_2014b}. It may also lead to a gradual
decline in manual skills of surgeons \cite{Steil_etal_2019}. Based
on the example of neurosurgical procedures, Sanniotis and Henneberg
describe this as a loss of a habit of routine that is necessary for
surgeons to acquire and maintain their skills. Adopting this habit,
however, as well as developing and maintaining ones’ skills to ones’
best abilities is understood to be a way to comply to the do-no-harm
principle \cite{Saniotis_Henneberg_2021}. As long as robots are available,
this would not have negative consequences, however, if circumstances
arise making manual intervention necessary, it would be detrimental
to human life \cite{Saniotis_Henneberg_2021}. 

Taking this line of argument one step further, Sullins warns that
phenomena of reverse adoption need to be closely observed. Reverse
adoption describes a common tendency in the introduction of technologies
in which social norms, practices and relations are changed and human
agents adapt to serve the functional needs of devices, instead of
devices adapting to serve users in fulfilling their goals \cite{Sullins_2014b}.
This subordination can be understood as a degradation of autonomy,
in which human users transform from those being able to set goals
to a means of maintaining technical function. It is, hence, noted
that especially on higher levels of functional autonomy the question
of RAS presents an analytic as well as an ethical challenge to define
acceptable pathways of human-machine cooperation in the operating
theater that preserve the human capacity to determine the ends. According
to Sullins, the first question is to define what it means to be autonomous
in this context while second is to clarify how this autonomy is compatible
with increasingly autonomous machines. The answer, as is argued, lies
in understanding cooperation as an ethical question that becomes increasingly
pressing with more complex devices. It should be explored by outlining
desirable forms of cooperation. Depicting forms of human-machine cooperation
as gradual continuum, positive notions of cooperation remain rather
unclear and do seldomly extend beyond ideas that connect to the theme
of meaningful control. On the negative side, instrumentalization described
as a degradation of human actors from those who set ends to those
who serve as means to ensure functioning of the machines \cite{Decker_2018b}
is understood to be unethical and a form of cooperation to be avoided.

With regard to the professional-patient-relationship, authors acknowledge
the importance of human-to-human contact as part of the surgical process
\cite{Sharkey_Sharkey_2012,Sharkey_Sharkey_2013,Narsinh_etal_2022,Giovagnoli_etal_2019,Nestor_Wilson_2019b,Datteri_2013b,Steil_etal_2019}.
The theme most discussed is whether distance between surgeons and
patients may affect mutual commitments \cite{Sullins_2014b}, potentially
weakening or disrupting trust and relationship between patients and
professionals (43,62). Objectification of patients, that is the reduction
of patients to objects of surgery rather than active participants,
should be avoided \cite{Nestor_Wilson_2019b,Sullins_2014b}, especially
in telesurgery. However, taking potential benefits into account as
well as the fact that potential phenomena of distancing may not extend
beyond the operating procedure itself, the authors in our dataset
seem to be less concerned about the impact on the professional-patient-relationship.

\section*{Discussion}

The main themes presented in our results testify to a broad range
of different and differing ethical considerations and perspectives on
RAS. In what follows, we comment on these findings in relation to
our two research questions: developing a systematic overview on the
ethical issues and to trace connections to technology and technological
properties within these debates. 
Regarding our first question, avoidance of harm, advancing patients’
interests, and maintaining control and responsibility over the process
of RAS seem to be commonly accepted primary ethical reference points.
Most authors in our dataset confine themselves to addressing only one of
these issues. Therefore, it remains unclear how these different principles
relate to or complement each other, or how they might be balanced
in case of conflict. Based on the generalized themes of this review,
we suggest that understanding each of these themes within the context
of a professional surgical ethos could help systematize the debates
and provide further insights.

McCullough et al. have argued that the surgical ethos sets particularly
high standards, drawing on the distinct circumstances of surgical
interventions \cite{McCullough_1998}, including the proximity of
surgeons and patients, the necessity to harm before healing, and
the patients’ acceptance of and surrender to these conditions. The authority
of surgeons over this process is understood as a necessary requirement.
However, from an ethical perspective it can only be granted conditionally
and is only justified if surgeons uphold the position of a trustworthy
fiduciary of patients’ interests while displaying and maintaining
the requisite skills, knowledge, and practical judgement as virtues
of the surgical profession \cite{McCullough_1998}. 

Against this background, our review indicates that RAS introduces
changing conditions for maintaining legitimate authority in operating
theater. For instance, unclear patient benefit or questionable trustworthiness
of evidence in regard to patient benefit do not only point to potential
research gaps, they also threaten the position of surgeons as moral
fiduciary, necessitating ethical considerations. Likewise, questions
of learning and questions of control point to surgeons techné and
knowledge as requirements controlling and using RAS devices properly.
In addition, this view shows that these standards might be riddled
with additional complexities and conflicting requirements that need
to be mitigated through proper judgement. For instance, while timely
adoption of RAS and proficiency through increasing case volume are
encouraged, there is also the call to uphold traditional skills through
constant practice, and to maintain the competence to step in if errors
occur. Mandating both as part of the robot surgeons ethos is demanding.
It requires further considerations to reconcile both duties to avoid
overburdening the surgical role with conflicting obligations. 

These considerations show that although RAS might rely on commonly
accepted ethical principles, satisfying the high standards of surgical
ethos requires adapting these principles to the circumstances of RAS.
This includes re-evaluating and re-orienting surgical duties and
professional ethos in light of new and emerging technological possibilities,
as well as identifying and mitigating value conflicts that arise with
these shifts. Looking to the future, we agree with authors in our
dataset who suggest a gradual shift that prescribes developing and
maintaining technical knowledge, competences and literacy in data
and digitization as part of the virtues for robotic surgeons, in addition
to their traditional skillset to meet these challenges .

With regard to our second research question, that is, how RAS and
its technical properties are addressed and linked to ethical dimensions
in the literature, we observe two differing perspectives. This becomes
apparent when comparing the analytic lenses through which devices
are examined in questions of responsibility and relational effects.
As noted above, a significant part of the included papers refers to
RAS in a conceptual language of “master-slave” devices, “slave systems”
or (enhanced) tools. These tools are described as means to a specific
end, with a low degree of autonomy, entirely dependent on the human
operator. With this passive role of the system, a clear distinction
between humans and machines is implied, since robotic surgical devices
are understood to merely translate hand movements into movements of
mechanical parts while fully respecting the surgeon’s intentions,
goals, and strategies. This perspective is equated with a “division
of labour” in which the human operator is in control by setting and
defining the rules, while the machine is confined to execute these.
Consequently, the devices are understood as an extension of surgeons’
physical and volitional capacities, which are completely at their
disposal \cite{vanderWaa_etal_2020}. In philosophy of technology,
this view or its derivative perspectives are often described as an
instrumental perspective in which technologies appear as means to
an end within human activity.

The second perspective, however, highlights the interaction, interplay,
and cooperation between human actors (e.g, the human surgeons or patients)
and the robotic devices. This view situates the machines as physical
entity with information processing capacities in the operating theater
\cite{Togni_etal_2021}. From this perspective, relationality is seen
as important feature \cite{Steil_etal_2019,Togni_etal_2021}. For
example, while the Da Vinci robot translates human movement into movement
of robotic instruments, it also provides surgeons with haptic feedback,
thus taking the position of a physical mediator between human surgeons
and patients. The robotic movements are intertwined with the surgeons’
as well as the patients’ bodies, integrating with the surgeon-patient-interaction.
This mediating position is understood as reshaping this interaction
in a novel way \cite{Togni_etal_2021}. With this shift, a distinct
active potential of robotic surgical devices in their role as mediators
based on their information processing capacities is noted. The devices
do not only transmit but actively transform what is and can be done
from the surgeons’ perspective. This view, hence, distances itself
from understanding RAS devised as passive machines \cite{Siciliano_Tamburrini_2019}.
Examples of this active potential include tremor suppression of surgeons’
hand movements or restricted movement path in newer systems \cite{Sharkey_Sharkey_2012},
as well as preprogrammed trajectories \cite{Saniotis_Henneberg_2021},
that execute or modify human actions \cite{Ficuciello_etal_2019}.
As, for example, Di Paolo concludes, RAS integrates a “new player”
in the operating theater \cite{DiPaolo_etal_2019}. This new player
breaks down the clear and embodied boundaries and sensations between
patients and surgeons through its bidirectional interaction \cite{Togni_etal_2021},
making division of labor and control less clear \cite{Steil_etal_2019},
and creating distinct and sometimes disruptive changes in human practices.

Proponents of the instrumental perspective are confident that using
RAS does not differ substantially from traditional interventions without
robots \cite{SiqueiraBatistaR._etal_2016}. With regard to responsibility
and professional ethos they, hence, argue that only very few new questions
besides potential harms and benefits arise, as compared to laparoscopic
or open procedures \cite{Smith_2013,SiqueiraBatistaR._etal_2016}.
The operating surgeon is seen as being fully in control and responsible
\cite{SiqueiraBatistaR._etal_2016,Smith_2013,Lirici_2022,Nestor_Wilson_2019b}.
Consequently, the ethical perspective that emerges from an instrumental
view is mostly based on considering trade-offs between improvement
of patient care and potential losses due to the means used in specific
conditions.

However, while this perspective seems to be intuitively plausible,
it does not account for effects and implications of using RAS that
are more difficult to quantify in terms of health outcomes or that
arise as a consequence of changing interactional patterns. In addition,
it does not seem to be fruitful to consider the effects of devices
gaining a more active role as described above. While a an instrumental
perspective may be suitable for devices that have low levels of functional
autonomy, it becomes inadequate as device activity increases. By contrast,
authors following the second perspective shift their focus on the
mediating position robotic devices occupy in the operating theater.
This view is, hence, tightly connected to questions of control and
professional obligations of surgeons, leading to considerations around
human-machine-interaction, inter-team relations, and relations between
patients and medical professionals. 

Based on these considerations, we understand the way devices are conceptualized
as an important precursor of the ethical analysis, as it provides
a focal lense through which RAS is addressed ethically. However, we
suggest that framing this as a mere ontological question that needs
to be settled to provide adequate ethical analysis would be an oversimplification.
Instead, such questions should be addressed at a conceptual level.
It can, hence, be understood as a question of so-called conceptual
ethics or conceptual engineering in approaching RAS \cite{Cappelen_Plunkett__2020}.
Conceptual ethics, broadly construed, is concerned with reflecting
on the choice of representational devices whenever that choice may
have non-conceptual consequences. The decision to use – or not use
– a particular concept can be based on a set of values appropriate
to the context and goals of use. With this in mind, we suggest understanding
the different perspectives in our dataset as offering complementary
focal lenses to the phenomena of RAS, which sometimes overlap without
being mutually exclusive and may be more or less suitable depending
on the context and goal in which they are put to use.

With regard to a future perspective on the analysis of RAS that aims
to anticipate upcoming ethical issues, questions of conceptual ethics
warrant further inspection and careful reflection. The current technological
development has allowed for – mostly implicitly – an instrumental
perspective to satisfy the conceptual conditions for a fruitful and comprehensive
analysis. While these conditions may be easier to satisfy with devices
with no or low functional autonomy, the likely shift towards more
active and autonomous devices calls for reflective caution with regard
to the presuppositions of future ethical inquiries to avoid missing
important questions. For instance, questions of control and responsibility
require to loosen the tight grip of thinking in means-end-schemes.
Providing answers in these matters requires rethinking conditions
of control and, hence, to consider the broader picture of interactional
patterns that allow to transfer control forth and back, or to step
in and take over control (for example in terms of feedback loops,
overriding mechanisms, timing windows for decision making etc.) \cite{Siciliano_Tamburrini_2019}.
This points to surgeons’ responsibilities in contributing to
developmental steps to create such conditions \cite{Collins_etal_2022}.

\subsection*{Strength and limitations}

Our paper provides a thorough and systematic review of the ethical
issues surrounding RAS, covering a wide range of literature from various
academic fields. With this, we integrate perspectives from philosophy,
medical ethics, philosophy of technology, medicine and computer science
to offer a comprehensive overview. We must concede, however, that
our findings come with at least two limitations. First, caution is
warranted with the criterion of operationalizing different levels
of functional autonomy of devices which prove more difficult than
expected and may have had an impact on inclusion and exclusion of
studies. Secondly, it is likely that our search strategy and selection
of sources may have resulted in a cultural bias in our dataset, limiting
the generalizability of the findings. Although we did not exclude
studies based on language, our dataset contained very few non-English
and non-Western records, making it more difficult to identify ethical
issues based on culturally different perspectives of RAS or medicine
and care in a more general sense. With this in mind we do not understand
our results to be exhaustive at this point. 

\section*{Acknowledgements}

We would like to thank Piotr Wilinski (University of Potsdam) for his support in 
saturation checking during our analysis. Fig. 2 was made by using icons from 
DinosoftLabs, dmitri13, Freepik, srip, Those Icons, Uniconlabs, Pixel perfect, 
Irfansusanto20, icon wind, Vector Stall, SeyfDesigner and Becris from 
flaticon.com 

\section*{Statements and declarations}
\subsection*{Funding}

This work was funded by the VolkswagenStiftung (9B 233) as part of the Digital 
Medical Ethics Network.

\subsection*{Competing interests}

The authors have no relevant financial or non-financial interests to disclose.

\subsection*{Author contributions}

RR and DP conceived the study. JH and RR designed the analysis. JH and RR wrote 
the protocol with contributions from DP and SPH. Preparation and data collection 
were performed by JH. Data extraction and analysis were performed by JH and 
SPH, supervised by RR and DP. The first draft of the manuscript was written by JH 
and all authors commented on previous versions of the manuscript. All authors 
read and approved the final manuscript.

\subsection*{Ethics approval}
The study did not involve human paticipants, animals, or their data or biological material. Ethics approval was not deemed necessary.

\subsection*{Consent to participate}
Not applicable

\subsection*{Consent to publish}
Not applicable

\bibliography{bibliography}

\end{document}